\documentclass[conference]{IEEEtran}
\IEEEoverridecommandlockouts
\usepackage{cite}
\usepackage{amsmath, amssymb, amsfonts}

\usepackage{graphicx}
\usepackage{textcomp}
\usepackage{xcolor}
\usepackage{algorithm, algpseudocode}

\def\BibTeX{{\rm B\kern-.05em{\sc i\kern-.025em b}\kern-.08em T\kern-.1667em\lower.7ex\hbox{E}\kern-.125emX}}

\begin{document}
  \title{Multi-agent Reinforcement Learning for Low-Carbon P2P Energy Trading among Self-Interested Microgrids\\

  \thanks{This work is partially supported by the A*Star SIMTech research fund.}
  \thanks{\textit{\textbf{Corresponding Author:}} Gaoxi Xiao.}
  }

  \author{\IEEEauthorblockN{1\textsuperscript{st} Junhao Ren} \IEEEauthorblockA{\textit{School of Electrical and Electronic Engineering} \\ \textit{Nanyang Technological University}\\ Singapore\\ junhao002@e.ntu.edu.sg}
  \and \IEEEauthorblockN{2\textsuperscript{nd} Honglin Gao} \IEEEauthorblockA{\textit{School of Electrical and Electronic Engineering} \\ \textit{Nanyang Technological University}\\ Singapore \\ honglin001@e.ntu.edu.sg}
  \and \IEEEauthorblockN{3\textsuperscript{rd} Lan Zhao} \IEEEauthorblockA{\textit{School of Electrical and Electronic Engineering} \\ \textit{Nanyang Technological University}\\ Singapore \\ zhao0468@e.ntu.edu.sg}
  \and \IEEEauthorblockN{4\textsuperscript{th} Qiyu Kang} \IEEEauthorblockA{\textit{School of Information Science and Technology} \\ \textit{University of Science and Technology of China}\\ Hefei, China\\ qiyukang@ustc.edu.cn}
  \and \IEEEauthorblockN{5\textsuperscript{th} Gaoxi Xiao} \IEEEauthorblockA{\textit{School of Electrical and Electronic Engineering} \\ \textit{Nanyang Technological University}\\ Singapore \\ egxxiao@ntu.edu.sg}
  \and \IEEEauthorblockN{6\textsuperscript{th} Yajuan Sun} \IEEEauthorblockA{\textit{Singapore Institute of Manufacturing Technology} \\ \textit{Agency for Science, Technology and Research}\\ Singapore \\ sun\_yajuan@simtech.a-star.edu.sg}
  }

  \maketitle

  \begin{abstract}
    Uncertainties in renewable generation and demand dynamics challenge day-ahead scheduling. To enhance renewable penetration and maintain intra-day balance, we develop a multi-agent reinforcement learning framework for self-interested microgrids participating in peer-to-peer (P2P) electricity trading. Each microgrid independently bids both price and quantity while optimizing its own profit via storage arbitrage under time-varying main-grid prices. A market-clearing mechanism coordinates trades and promotes incentive compatibility. Simulations show that the learned bidding policy improves renewable utilization and reduces reliance on high-carbon electricity, while increasing community-level economic welfare, delivering a win–win in emission reduction and local prosperity.
  \end{abstract}

  \section{Introduction}
  Carbon emissions, the primary driver of the greenhouse effect, have severely constrained global sustainable development. To address this challenge, the Paris Agreement \cite{horowitz2016paris} set the goal of achieving net-zero emissions by 2050. Among various mitigation strategies, integrating renewable energy into conventional power networks has become a key approach. Renewable sources are expected to account for $25\%$–$41\%$ of global electricity generation by 2040 \cite{dengPowerSystemPlanning2020}.

  In practice, most microgrids procure the majority of their electricity through the day-ahead (DA) market based on forecasts of demand and renewable generation. However, due to the inherent uncertainty of renewable resources and load dynamics, significant deviations between day-ahead schedules and real-time realizations are inevitable though equiped with energy storage system can relieve these fluctuations \cite{kwonOptimalDayAheadPower2017}. As a result, microgrids frequently experience unexpected surpluses or deficits during operation, which cannot be efficiently resolved by the day-ahead market alone.

  To address this issue, intra-day peer-to-peer (P2P) electricity trading has been widely advocated as a flexible mechanism to rebalance local supply and demand among neighboring microgrids \cite{morstynUsingPeertopeerEnergytrading2018}. Nevertheless, designing effective P2P trading mechanisms remains challenging due to the decentralized market structure and  inherently complex, stochastic, and highly nonconvex characteristics.
  Under such conditions, microgrids are required to make sequential and adaptive bidding and storage control decisions based on limited local information and repeated market interactions. This naturally leads to a decentralized sequential decision-making problem, for which multi-agent reinforcement learning (MARL) provides a principled and scalable solution framework.

  MARL enables microgrids to learn adaptive decision policies directly from repeated market interactions, without relying on accurate system models or complete information, thereby endowing them with the ability to respond autonomously to evolving market conditions and endogenous strategic dynamics \cite{qiuMultiagentReinforcementLearning2021}. This capability is particularly essential for realizing self-organized, low-carbon, and intelligent P2P energy trading systems.

  In this paper, we propose a MARL-based double auction bidding strategy for microgrids to maximize their own interests in a P2P electricity market. With the help of an effective market clearing mechanism, the learned bidding strategy can protect individual interests while also reducing overall carbon emissions in the community and ensure social welfare. Our contributions are concluded as follows.
  \begin{itemize}
    \item We formulate the multi-microgrid P2P market as a decentralized partially observable Markov decision process (DEC-POMDP) and solve it under a centralized training with decentralized execution (CTDE) paradigm to mitigate day-ahead dispatch errors. After P2P clearing, any remaining surplus can be actively fed into the main grid at the feed-in tariff (FIT) without requiring storage to be full.
    \item We design a multi-agent proximal policy optimization (MAPPO) framework that incorporates a Long Short-Term Memory (LSTM) network to extract multi-scale temporal features, thereby improving bidding decisions and storage management. In addition, effective market-clearing mechanisms are integrated to ensure system-level efficiency while preserving individual economic incentives. Unlike prior studies, each microgrid in the proposed framework simultaneously decides price, quantity, and the storage parameter to maximize its own profit in this electricity market.
  \end{itemize}
  The remainder of the paper is organized as follows. Section II reviews existing methods and technologies employed in the P2P market of multi-microgrid systems. Section III presents the overall system formulation, including the structure of the multi-microgrid systems and the electricity market framework. Section IV describes the proposed MARL-based model for the P2P market. Section V provides simulation results and discussions. Section VI concludes the paper and outlines potential directions for future research.
  \section{Related Works}
  Existing studies on trading problems in multi-microgrid systems mainly rely on two methodological paradigms: model-based optimization and model-free optimization (specifically, MARL).

  Model-based optimization refers to frameworks in which the objective functions, constraints, and system dynamics are explicitly known or can be analytically modeled, allowing optimal decisions to be directly derived from the model. In \cite{liRiskaverseEnergyTrading2017}, a two-stage stochastic game model was proposed to determine the optimal energy trading strategy under supply and demand uncertainties. A distributed algorithm was further developed to obtain the stochastic Nash equilibrium using the sample average approximation technique. A stochastic cartel game was introduced in \cite{wangStochasticCooperativeBidding2022} to address cooperative bidding strategies among multiple microgrids in P2P energy transactions considering renewable generation and energy storage. A collaborative optimization problem for capacity planning of distributed generation units and P2P trading was investigated in \cite{wangRiskaverseStochasticCapacity2024}, where a risk-averse stochastic programming framework and Nash bargaining approach were adopted to mitigate renewable energy uncertainties and ensure fairness in trading. In addition, a low-carbon P2P trading model based on a master–slave nested mixed game was proposed in \cite{liangMultiagentLowcarbonOptimal2024}, which explicitly incorporated carbon emission flows as constraints in the game formulation.

  Model-free optimization, on the other hand, enables agents to learn their individual trading strategies and energy schedules from historical data to maximize operational profitability without requiring explicit knowledge of system dynamics. In \cite{qiuMultiagentReinforcementLearning2021}, a double-sided auction-based P2P electricity market was formulated as a MARL problem, and the DA-MADDPG algorithm was proposed to maximize the profits of prosumers in a dynamic electricity market. A two-level MARL framework was developed in \cite{mayMultiagentReinforcementLearning2023} to design dynamic pricing policies that facilitate efficient on-site energy trading while supporting decarbonization and grid security objectives. In \cite{yangMultistageStochasticDispatching2024}, a multi-stage dispatching method incorporating day-ahead and intra-day scheduling was proposed to address uncertainties in electricity–hydrogen integrated energy systems. Moreover, \cite{chenCombinedCarbonCapture2024} introduced a multi-microgrid framework combining carbon capture and utilization technologies with P2P energy trading, trained using the MAPPO algorithm to reduce both cost and carbon emissions. A recurrent neural network (RNN)-based MAPPO algorithm was further presented in \cite{zhouJointEnergyCarbon2024} to learn one-to-one clearing policies in the coupled energy and carbon trading markets.

  \section{Problem Statement}
\begin{figure}
    \centering
    \includegraphics[height=0.25\textwidth]{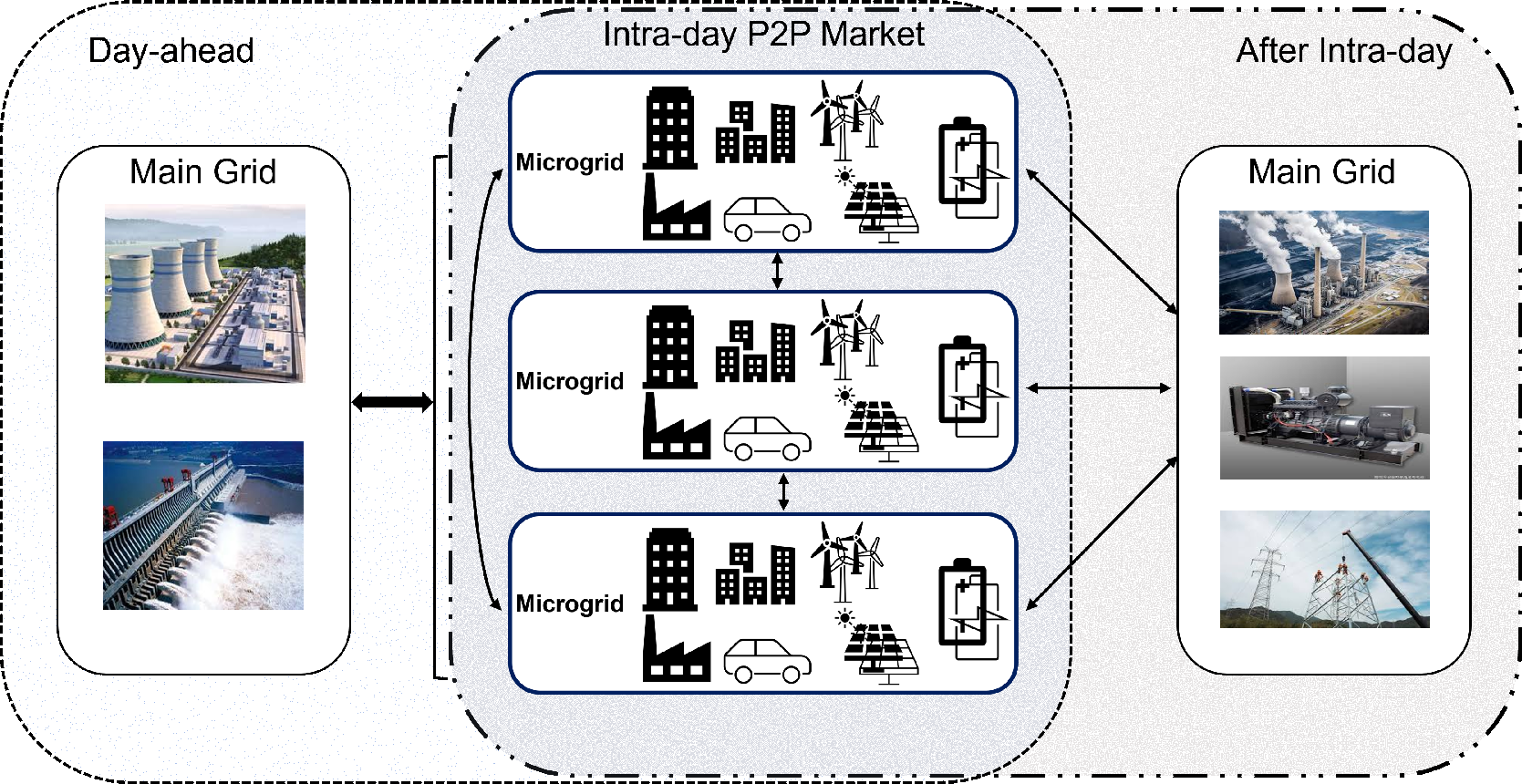}
    \caption{The two-stage electricity market framework.}
    \label{fig:market_framework}
  \end{figure}

  In this paper, we investigate a network of interconnected microgrids, as illustrated
  in Fig.~\ref{fig:market_framework}. The network consists of a main grid and $N$
  distributed microgrids, each representing different types of regions—such as
  residential, industrial, and commercial areas. The main grid comprises heterogeneous
  generation resources, and electricity procurement across different market stages is
  considered for assessing both community welfare (economic payoff) and societal
  welfare (carbon emissions). The heterogeneity of these resources implies that the
  main grid can supply electricity with varying carbon emission intensities,
  which will be detailed in Section ~\ref{subsection: eletricitymarket}. Each
  microgrid is equipped with electrical loads, photovoltaic (PV) generation, and
  electricity storage (ES), which provide operational flexibility and resilience
  to emergency. For tractability, this work focuses on economic performance and carbon
  accounting, while the validation of physical power-flow constraints (e.g., line
  flows and AC power-flow equations) is omitted.

  \subsection{Microgrid System}
  For the microgrid $i$, let $L_{i,t}$ denote its electricity demand and $G_{i,t}$
  denote its photovoltaic renewable electricity output at hour $t$. In
  operational process, microgrids prioritize using ES to balance their own deficits
  and surplus after P2P trading. This means ES acts as a power supply to discharge
  electricity if generation and day-ahead purchases are insufficient, otherwise
  it is the load to charge the oversupply. Therefore, the state of charge (SoC)
  of ES for each microgrid $i$ at hour $t$ is given by:
  \begin{equation}
    \label{eq: SoC}E_{i, t+1}=
    \begin{cases}
      E_{i,t}+ \dfrac{\beta_{\mathrm{chr}} T^{\text{ES}}_{i,t} \Delta t}{E^i}, & \text{if }T^{\text{ES}}_{i,t}\geq 0, \\
      E_{i,t}+ \dfrac{T^{\text{ES}}_{i,t} \Delta t}{\beta_{\mathrm{dis}} E^i}, & \text{if }T^{ES}_{i,t}< 0,
    \end{cases}
  \end{equation}
  where $T^{ES}_{i,t}$ and $E^{i}$ denote the electricity power and current capacity
  of ES for microgrid $i$ at hour $t$, respectively. The initial capacity of microgrid
  $i$ is denoted by $E^{i}_{0}$. $\beta_{\mathrm{chr}}$ and
  $\beta_{\mathrm{dis}}$ refer to the charge and discharge efficiency of ES for
  microgrid $i$. The electricity power $T^{ES}_{i,t}$ and SoC $E_{i, t+1}$ should
  satisfy the following inequalities:
  \begin{equation}
    \label{eq: ESconstraints}
    \begin{aligned}
      \underline{T}^{\text{ES}}_{i} & \le T^{\text{ES}}_{i,t}\le \overline{T}^{\text{ES}}_{i,t} \\
      E_{i, \min}            & \le E_{i, t}\le \alpha_{e, i}E_{i, \max}
    \end{aligned}
  \end{equation}
  where $\underline{T}^{\text{ES}}_{i}$ and $\overline{T}^{\text{ES}}_{i,t}$ denote the hourly
  minimum and maximum electricity power of ES for microgrid $i$. $E_{i, \min}$,
  $E_{i, \max}$ and $\alpha_{e}$ are the minimum, maximum storage level, and
  control parameters of ES, respectively. Without loss of generality, we allows the storage device to simultaneously absorb and supply power within the same trading interval, as commonly observed in practical multi-port or bidirectional converter-based energy storage systems.

  \subsection{Electricity Market Framework}
  \label{subsection: eletricitymarket} In this work, we consider a two-stage
  electricity market for microgrid participants, consisting of a day-ahead (DA)
  stage and an intra-day stage (see Fig. \ref{fig:market_framework}). The day-ahead stage is
  used to provide the majority of the electricity of microgrids based on forecasts
  of demand and renewable generation, while the intra-day P2P market is to rebalance
  surpluses and deficits due to uncertainties of forecasts.

  \subsubsection{DA market mechanism}
  The day-ahead market is widely recognized as an efficient clearing mechanism
  for electricity and other commodities. In this stage, the main grid announces
  three distinct price signals to microgrids: the day-ahead price $p^{\text{G}}
  _{\text{da}}$, the emergency price $p^{\text{G}}_{\text{e}}$, and the
  FIT $p^{\text{G}}_{\text{f}}$. Since emergency electricity is typically
  generated by fuel-powered or diesel units with high carbon emissions, and excessive
  feed-in electricity from distributed sources may threaten grid stability, it is
  reasonable to assume that
  \begin{equation}
    p^{\text{G}}_{\text{f}}\le p^{\text{G}}_{\text{da}}\le p^{\text{G}}_{\text{e}}
    ,
  \end{equation}
  which penalizes both emergency electricity usage and excessive feed-in to the
  main grid. Without loss of generality, it is assumed that $p^{\text{G}}_{\text{f}}$ is time-invariant while $p^{\text{G}}_{\text{e}}$ is time-variant. By differentiating these prices, the main grid can ensure reliable baseline generation while simultaneously encouraging the integration of renewable resources. Each microgrid $i$ then forecasts its hourly demand and renewable generation for the following day. However, due to the limited capacity of local
  renewable units and inevitable forecast errors, some microgrids may still experience electricity deficits that must be compensated by purchasing electricity from the main grid in the day-ahead market. The day-ahead procurement policy of microgrid $i$ at hour $t$ is modeled as
  \begin{equation}
    \label{eq:dayaheadpolicy}q^{\mathrm{da}}_{i,t}= \max \{0, \alpha_{i} (\bar L_{i,t}
    - \bar G_{i,t})\},
  \end{equation}
  where $\bar L_{i, t}\ge 0$ and $\bar G_{i, t}\ge 0$ denote the forecast load and
  local renewable generation of microgrid $i$ at hour $t \in [0, 24)$. Positive
  values of $q^{\mathrm{da}}_{i,t}$ indicate purchases from the main grid; sales
  to the main grid are excluded from the day-ahead stage and are addressed after the intra-day
  P2P market. $\alpha_{i}$ denotes the procurement policy factor of microgrid
  $i$ in the day-ahead stage. In this paper, the day-ahead procurement policy factor
  $\alpha_{i}$ of the microgrid $i$ remains fixed, and therefore we focus on using the
  intra-day P2P market to balance residual surpluses and deficits to improve
 the' economic outcomes of the participants.

  \subsection{P2P market mechanism}
  Due to uncertainties on forecasts of generation and loads, a double
  auction clearing mechanism is thereby needed to promote internal balance among these microgrids and the integration of renewable energy. In this stage, microgrids should make
  the decision to buy/sell how much electricity and at what price given the observed
  information.

  At each time step in the P2P market, three sequential events occur for each microgrid $i$: (1) submission of a quotation $(p_{i}, q_{i})$, (2) matching of quotations through the P2P market mechanism, (3) electricity exchange and payment settlement. In the quotation $(p_{i}, q_{i})$, $ p^G_f \le p_{i} \le p^G_e$ represents the unit price at which the microgrid intends to buy/sell electricity, and $0 \le q_{i} \le \bar q_{i, t}$ denotes the amount of electricity it wishes to trade while $\bar q_{i}$ represents the current maximum demand/supply electricity of microgrid $i$. This enables microgrids to offer more diverse quotations in transactions, thereby maximizing their own benefits. This bidding strategy allows microgrids to submit flexible and diverse offers in the market, thereby optimizing their own operational benefits and reflecting self-interested trading behavior.

  In the matching stage, the multi-round double auction clearing (MRDAC) mechanism proposed in \cite{haggiMultiRoundDoubleAuctionEnabled2021} is employed to efficiently clear the bids. In the MRDAC mechanism, the auctioneer first collects all quotations, where $n_b$ quotations are submitted by buyers and $n_s$ quotations are submitted by sellers. The quotations from sellers are then arranged in ascending order of their offered prices, while those from buyers are arranged in descending order of their bid prices. For instance, suppose the auctioneer receives quotations from three sellers: $(6, 2)$, $(1, 3)$, and $(4, 2)$, where the first element denotes the unit price and the second the quantity. The matching priority for these seller quotations is given by $(1, 3) > (4, 2) > (6, 2)$, meaning that sellers offering lower prices are prioritized. A similar rule applies inversely to the buyer side.

  After ranking the quotations from buyers and sellers, the auctioneer matches the highest-priority seller $i$ with the highest-priority buyer $j$. The transaction is then settled at the \textit{average price}
  \begin{equation}
    p_{i,j} = \frac{p_i + p_j}{2} = p_{j,i},
  \end{equation}
  and the \textit{traded quantity} is determined by
  \begin{equation}
    q_{i,j} = \min(q_i, q_j) = q_{j,i}.
  \end{equation}
  If a seller only sells part of its quoted quantity, the remaining quota is moved to the end of the sellers’ list to await subsequent matching rounds, while the next seller in priority begins trading with the next buyer. The same procedure applies symmetrically to buyers who do not fulfill their desired purchase quantities.

  During the matching process, a transaction between seller $i$ and buyer $j$ can be settled only if $p_i \le p_j$. Otherwise, one of the two participants is randomly moved to the end of its respective list to await the next matching round. In addition, a seller (or buyer) is removed from the waiting list once its quoted quantity has been fully allocated, or when its unit price becomes higher (or lower) than all remaining buyers (or sellers). The matching process terminates when there are no remaining quotations in the waiting list for either buyers or sellers, or for both.

  At the end of the matching process, all microgrids with electricity deficits must purchase emergency electricity from the main grid at the price $p^{\text{G}}_{\text{e}}$, while those with electricity surpluses sell their excess energy to the main grid at the price $p^{\text{G}}_{\text{f}}$. Overall, the MRDAC mechanism with an average-price settlement effectively enhances the economic incentives for small-scale microgrids to participate in the P2P market, as it is generally more cost-efficient than trading directly with the main grid after the P2P stage.

  \section{MARL-based Bidding Strategy in Intra-day P2P Market}
  In this section, the intra-day P2P trading problem is formulated as a DEC-POMDP, which is typically solved using MARL algorithms. In this work, we adopt a CTDE paradigm and propose a MAPPO framework integrated with a LSTM architecture (MMAPPO), which enables each microgrid to learn an autonomous bidding policy from repeated market interactions.

  \subsection{Model of bidding strategy for microgrid $i$}
  In the intra-day P2P market, each microgrid $i$ aims to maximize its own economic benefit by optimizing its bidding strategy subject to operational and market constraints. The bidding strategy model for microgrid $i$ can be formulated as follows:
  \begin{align}
    \max_{p_{i,t},\, q_{i,t},\, \alpha_{e, i, t}} 
    \sum_{t=1}^{T} & \left( P^{\text{G}}_{i,t} + P^{\text{p2p}}_{i,t} \right)  \qquad  \label{eq:obj}\\
    \text{\quad s.t.} \quad
     L_{i,t} + T^{\text{ES}}_{i,t} + q^{\text{fit}}_{i,t} + q^{\text{s}}_{i,t} 
      = &G_{i,t} + q^{\text{da}}_{i,t} + q^{\text{b}}_{i,t} + q^{\text{e}}_{i,t}, \label{eq:balance}\\
     p^{\text{G}}_{\text{f}} \le  p_{i,t} &\le p^{\text{G}}_{\text{e}}, \label{eq:price}\\
     -\bar{q}_{i,t,\text{s}} \le  q_{i,t} &\le \bar{q}_{i,t,\text{b}}, \label{eq:quantity}\\
     0 \le  \alpha_{e, i, t} & \le 1, \label{eq:alpha}\\
      Eq.~\eqref{eq: ESconstraints}.& \nonumber
  \end{align}
  Here, $P^{\text{G}}_{i,t}$ denotes the operational profit of microgrid $i$ related to the main grid, and $P^{\text{p2p}}_{i,t}$ represents the profit obtained from P2P trading. The variables $q^{\text{fit}}_{i,t}$, $q^{\text{e}}_{i,t}$, $q^{\text{s}}_{i,t}$, and $q^{\text{b}}_{i,t}$ correspond to the quantities of feed-in electricity to the main grid, emergency electricity purchased from the main grid, electricity sold, and electricity bought in the P2P market, respectively. The parameters $\bar{q}_{i,t,\text{s}}$ and $\bar{q}_{i,t,\text{b}}$ denote the maximum supply and demand capacities of microgrid $i$ when acting as a seller or buyer, respectively. In addition, for each microgrid $i$, $P^{\text{G}}_{i,t}$ only considers the profit obtained from feed-in and emergency transactions, since the expected day-ahead expense remains constant when $\alpha_i$ is fixed. Therefore,
  \begin{equation}
    P^{\text{G}}_{i,t} = p^{\text{G}}_{\text{f}} q^{\text{fit}}_{i,t} 
    - p^{\text{G}}_{\text{e}} q^{\text{e}}_{i,t}.
  \end{equation}
  Similarly, the profit of microgrid $i$ from P2P trading is given by
  \begin{equation}
    P^{\text{p2p}}_{i,t} = 
    \sum_{j \in n_b} p_{i,j,t} q_{i,j,t} 
    - \sum_{j \in n_s} p_{i,j,t} q_{i,j,t},
  \end{equation}
  where $n_b$ and $n_s$ denote the sets of matched buyers and sellers, respectively, and $n_b + n_s \le N$.

  \subsection{Model of DEC-POMDP}
  The P2P trading problem can be formulated as a DEC-POMDP. A DEC-POMDP is defined by the tuple 
  \[
  \left( \mathcal{N}, \mathcal{S}, \mathcal{A}^i, \mathcal{O}^i, \mathcal{R}^i, \mathcal{P}, \gamma \right),
  \]
  where $\mathcal{N}$, $\mathcal{S}$, and $\mathcal{P}$ denote the set of agents, the global state space of the environment, and the state transition function, respectively. The components $\mathcal{A}^i$, $\mathcal{O}^i: \mathcal{S} \times \mathcal{A} \times \mathcal{S} \to [0,1]$, and $\mathcal{R}^i: \mathcal{S} \times \mathcal{A} \to \mathbb{R}$ represent the action space, the observation space, and the reward function for agent $i \in \mathcal{N}$, respectively. The discount factor $\gamma \in (0,1)$ accounts for the agent's preference for immediate versus future rewards. 

  \subsubsection{Agents} Each agent $i$ corresponds to a distinct microgrid in the system, where all agents collectively participate in the P2P trading process. 

  \subsubsection{Observation} At time step $t$, the partial observation available to agent $i$ is defined as
  \begin{equation}
    \mathcal{O}^i_t = \{m_t, E_{i,t}, h_{i,t}, t\},
  \end{equation}
  where $m_t$ represents the embedded market information reflecting the overall supply–demand condition in the P2P market; 
  $E_{i,t}$ denotes the current energy storage (ES) capacity of microgrid $i$; 
  and $h_{i,t}=\{q^{\text{da}}_{i,z}, \bar{L}_{i,z}, \bar{G}_{i,z}, p^G_{e,z}\}_{z=t-\delta_1}^{t+\delta_2}$ 
  is the noisy temporal observation embedding that includes the day-ahead procurement quantity, 
  forecasted load, forecasted renewable generation, and the emergency electricity price within the time interval $(t-\delta_1, t+\delta_2)$. 
  The temporal window allows each agent to incorporate both historical and predictive information into its local decision-making process. 

  \subsubsection{State} The global state of the environment at time $t$ is denoted by 
  $\mathcal{S}_t = \{{s}_{1,t}, {s}_{2,t}, \dots, {s}_{N,t}\}$, which consists of all possible environment states determined by the agents' actions and rewards.

  \subsubsection{Action} At each time step $t$, agent $i$ selects an action 
  \[
  \mathcal{A}^i_t = \{p_{i,t}, q_{i,t}, \alpha_{e,i,t}\},
  \]
  where $p_{i,t}$ and $q_{i,t}$ represent the bidding price and quantity in the P2P market, respectively, and $\alpha_{e,i,t}$ denotes current ES control parameter of microgrid $i$. 
  The joint action of all agents is $\mathcal{A}_t = \{\mathcal{A}^1_t, \dots, \mathcal{A}^N_t\}$.

  \subsubsection{Reward} After the environment transitions from $\mathcal{S}_t$ to $\mathcal{S}_{t+1}$ due to the joint action $\mathcal{A}_t$, 
  agent $i$ receives an immediate reward $\mathcal{R}^i_t$ defined as
  \begin{equation}
    \mathcal{R}^i_t = P^{\text{G}}_{i,t} + P^{\text{p2p}}_{i,t}.
  \end{equation}
  The proposed design seeks to maximize operational profitability through the integration of renewable energy sources while simultaneously reducing reliance on electricity resources characterized by high carbon emissions.

  \subsection{MMAPPO-driven intra-day P2P market}
  In this work, the MMAPPO algorithm is presented to address P2P trading problem among $N$ microgrids. The MMAPPO framework consists of $N$ decentralized policy networks, denoted by $\{\pi_{\theta_i}\}_{i=1}^{N}$, and a centralized critic network $V_{\phi}$, which are used to approximate the policy and value functions of the microgrids, respectively. 
  
  To enhance temporal representation, an LSTM module is incorporated for sequential observation embedding, while a sinusoidal periodic encoder is employed to capture daily cyclical patterns. The overall training process of MMAPPO is summarized in Algorithm~\ref{alg:mmappo}. For further architectural and implementation details of the policy and critic networks, readers are referred to \cite{yu_surprising_2022}.
  \begin{algorithm}[!b]
    \caption{MMAPPO for P2P Microgrids}
    \label{alg:mmappo}
    \begin{algorithmic}[1]
    \State Initialize simulation parameters.
    \State Initialize policy networks $\{\pi_{\theta_i}\}_{i=1}^N$, critic network $V_\phi$, buffers $\{\mathcal{D}_i\}_{i=1}^N$.
    \For{$episode=1$ to $N_{\mathrm{ep}}$}
    \State Reset the environment and Clear all buffers.
      \For{$t=1$ to $T$}
          \State Actions sampling $\{\mathcal{A}^i_t \sim \pi_{\theta_i}(\cdot|\mathcal{O}^i_t)\}_{i=1}^N$.
          \State Environment steps $\to \mathcal{S}_{t+1}$, $\{\mathcal{R}^i_t\}_{i=1}^N$.
          \State Store $\{o^i_t,a^i_t,\mathcal{R}^i_t,o^i_{t+1},s_t,s_{t+1}\} \to \mathcal{D}_i$.
      \EndFor
      \For{$epoch=1$ to $K$}
          \For{each mini-batch $\mathcal{B}$ from buffers}
              \State Normalized advantages $\{A^i_t\}_{i=1}^N \gets V_\phi(s_t)$.
              \State Update each $\theta_i$ via per-agent policy loss.
              \State Update shared $\phi$ via critic loss.
            \EndFor
        \EndFor
    \EndFor
    \end{algorithmic}
  \end{algorithm}
  \section{Numerical Simulation}
  \subsection{Environment Setup}
  We consider an electricity market comprising four microgrids interconnected through a shared main grid. The parameters of these four microgrids are listed in Table~\ref{tab:grid-parameters}. In the simulation, it is assumed that there is no energy loss during charging and discharging processes, i.e., $\beta_{\text{dis}} = \beta_{\text{chr}} = 1$.  The parameters of the observable time window for each microgrid are set as $\delta_1 = 1$ and $\delta_2 = 6$. Furthermore, the daily renewable generation and electricity load profiles of the four microgrids are sampled from the normalized data of four Australian residential households reported in \cite{ratnam2017Residential}. The feed-in tariff (FIT) price of the main grid is fixed at $2~\$/\text{kWh}$, while the emergency price varies dynamically over time within the range of $[15, 35]~\$/\text{kWh}$.
  \begin{table}[t]
    \caption{Simulation Parameters for the 4 Microgrids}
    \begin{center}
      \begin{tabular}{|c|c|c|c|c|}
        \hline
        \textbf{Parameter} & \textbf{Grid 1} & \textbf{Grid 2} & \textbf{Grid 3} & \textbf{Grid 4} \\
        \hline
        $L_{i}^{\text{max}}$ (kWh) & 25 & 6 & 40 & 5 \\
        \hline
        $G_{i}^{\text{max}}$ (kWh)& 5 & 7 & 10 & 15 \\
        \hline
        $E_{\text{max}}$ (kWh)& 8 & 15 & 15 & 30 \\
        \hline
         $\overline{T}^{\text{ES}}_{i}$ (kW)& 4 & 5 & 8 & 10 \\
        \hline
         $-\underline{T}^{\text{ES}}_{i}$ (kW)& 4 & 5 & 8 & 10 \\
        \hline
        $E_0$ (kWh)& 0 & 2 & 0 & 20 \\
        \hline
      \end{tabular}
      \label{tab:grid-parameters}
    \end{center}
  \end{table}
  
  \subsection{Baselines} 
  We compare the proposed framework against three existing multi-agent reinforcement learning (MARL) algorithms: 
  (1) \textbf{MIPPO} --- an improved independent PPO algorithm with an LSTM-based architecture; (2) \textbf{MAPPO-one} --- a MAPPO variant that employs an LSTM encoder but shares a single global critic across all agents; and (3) \textbf{MAPPO-s} --- a simplified MAPPO-one configuration with a smaller critic network.  In addition, we evaluate the performance of different market-clearing mechanisms (\textbf{Greedy} and \textbf{VDA} \cite{zhaoComparisonsAuctionDesigns2023}) against the existing MRDAC mechanism to assess the advantages of the proposed framework.
  \begin{figure}
      \centering
      \includegraphics[width=0.8\linewidth]{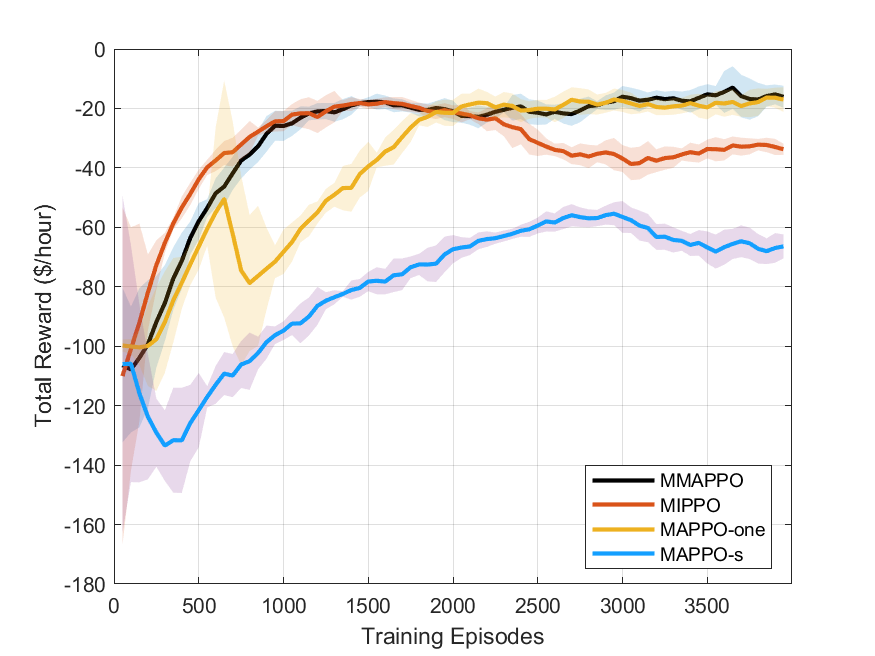}
      \caption{Training curves of different MARL algorithm $+$ MRDA.}
      \label{fig:baseline1}
  \end{figure}

  \subsection{Results \& Discussions}
    As shown in the training results in Fig.~\ref{fig:baseline1}, the proposed MMAPPO algorithm achieves the highest total reward (approximately $-15$ on average) and demonstrates the fastest convergence within about 1000 episodes, outperforming MAPPO-s and MIPPO by roughly $20 \$ $ per hour in the steady stage. In contrast, MIPPO quickly improves at the early stage but soon becomes trapped in a local optimum, while MAPPO-one converges more slowly, requiring around 2000 episodes. The performance of MAPPO-s declines significantly due to the limited capacity of its critic network.

    \begin{table}[htbp]
    \caption{Daily Performance Comparison of MMAPPO under Different Market Mechanisms}
    \centering
    \setlength{\tabcolsep}{5pt}
    \renewcommand{\arraystretch}{1.15}
    \begin{tabular}{lccc}
    \hline
    \textbf{Metric} & \textbf{MRDA} & \textbf{VDA} & \textbf{ Greedy} \\
    \hline
    Total Profit (\$)             & $-123.81$ & $-185.53$ & $-221.02$ \\
    Emergency Purchase (kWh)      & $30.33$   & $34.00$   & $34.37$   \\
    Total FIT (kWh)             & $18.95$   & $32.24$   & $43.74$   \\
    P2P Trade Volume (kWh)               & $10.58$   & $9.96$    & $9.55$    \\
    Average SoC (kWh)         & $16.56$   & $16.69$   & $16.60$   \\
    \hline
    \end{tabular}
    \label{tab:baseline2}
    \end{table}
     For performance comparison of $4$ microgrids under different market-clearing mechanisms, MMAPPO was trained under each mechanism for approximately $5000$ episodes and evaluated over $960$ testing days. The average results are summarized in Table~\ref{tab:baseline2}. The table shows that the MRDAC mechanism under MMAPPO achieves the best economic and operational performance among the three mechanisms. Specifically, MRDAC yields the highest total profit (improving by $50\%$ over VDA and $78\%$ over Greedy), and the lowest emergency purchase (reducing by $12\%$ and $13\%$ compared with VDA and Greedy, respectively). Moreover, the MRDAC mechanism significantly increases the trading volume in the P2P market. These results demonstrate that MRDA, when integrated with MMAPPO, effectively enhances renewable energy utilization and maintains strong individual trading incentives.

  \section{Conclusion}
  In this paper, we proposed a double-auction-based P2P trading framework for multi-microgrid systems using the MMAPPO algorithm. The trading problem under the MRDAC mechanism was formulated as a DEC-POMDP, enabling each microgrid to make autonomous and coordinated decisions under uncertainty. By integrating MAPPO with an LSTM architecture, the framework achieved efficient and reliable decision-making through multi-scale feature extraction. Simulation results verified that the proposed MARL-based mechanism enhances the adaptability and intelligence of distributed energy management. Future work will extend this study to optimize day-ahead procurement policy and address uncertainty caused by renewable generation faults.

  \bibliographystyle{IEEEtran}
  \bibliography{IEEEabrv,ref}
\end{document}